# BUILDING ACTIVE AND COLLABORATIVE LEARNING ENVIRONMENT IN INTRODUCTORY PHYSICS COURSE OF FACULTY OF TECHNOLOGY EDUCATION

*Nguyen Hoai Nam, PhD., namnh@hnue.edu.vn,– Faculty of Technology Education, Hanoi University of Education, Hanoi, Vietnam*

**Abstract.** Model of active and collaborative learning applied in training specific subject makes clear advantage due to the goals of knowledge, enhanced activeness, skills that students got to develop successful future job. Studying and applying the model to build a learning environment in the Introductory Physics course with the support of ICT, especially social network and eLearning system of HNUE.

## 1. Introduction

Vietnam Education is on the first step of innovation: changing from content training focus in developing competency-based learner training. To make a success, both content and program must be redesigned from teacher-oriented to learner-oriented. The teacher becomes an instructor to help the student in learning.

HNUE is a key university of training teacher, therefore it is asked to study solutions to enhance quality of future instructor to meet education reform requirement.

Researching solution is not a task of Methodology Unit but teaching member in any subject of higher education, because the Freshman learns knowledge and training business from the instructor.

## 2. Content
### 2.1. General issues of the Freshman of Faculty of Technology Education

The Freshman has just left from a hard entrance examination. There are several disadvantages factors to learner included playing desires or uneasy feeling due to a temporary learning mental process of who not be in the proper environment. Nevertheless, lasting the duration for accepting the $2^{nd}$ registration affects disadvantageously on student, especially who begin later.

The Freshman gets difficult and nervous of a new learning way which needed to pay more attention for self-learning. They also have to adapt to a new environment with many challenges because that is the first time apart from family.

According to credit training tree, the Introductory Physics is a common and a prerequisite subject like Mathematics, Introductory Informatics for the $1^{st}$ year student before registering the next business. [13] Most of introductory physics contents mentioned in school with easier requirement and experience with practicing in the progress training for university entrance of leaners make them subjective feel. However, the content of the $1^{st}$ part of the introductory physics is most related to mechanics and heat, which they had studied in the first year of school so it makes some trouble to remember. Many students come from rural and mountainous region therefore the ICT (Information and Communication Technology) skills are limited.

But the student can adapt quickly to changing and eager to learn a new thing which reveals a suggestion to exploit this feature of learner to enhance learning effectiveness.

**2.2. Active and Collaborative Learning Model (ACLM)**

Active learning based on constructivism theory with two strands: cognitive constructivism and social constructivism. Several constructivists ideas have been used to inform adult education, whereas pedagogy applies to the education of children. There are different in emphasis, but they also share many common perspectives about teaching and learning. [14] With the support of ICT, the collaboration of learner is essential [2], [12]. Different from the traditional model in which listening and writing passively, students are encouraged to join in activities with the active learning model. [1] They actively study learning materials such as documents, textbooks... or searching necessary contents with the help of instructor to get knowledge and skills. [10] Student can work independently, in pair or group and be encouraged to present and defend result in group or class. Many researchers agree that most of learners feel excited and have advantage result due to that work. [11] [4] [6]

In the learning model, students play a major role to find knowledge. They practice skills to solve problems, working in a team, presenting, asking, answering… Students also actively join in evaluating process included self-evaluation, group evaluation and external evaluation. By the way, they absorb knowledge, correct wrong information and learn more experiences from learning partner (classmate…) and teacher (instructor).

As an instructor, the teacher also plays an important role because giving suggestion to help student solving problem efficiently. The instructor also analyses, evaluates and corrects the wrong in cognitive process and inexactly self-evaluating, peer-evaluating of students.

The efficiency of learning activities improved with the support of ICT as various rich multimedia contents, tools for searching, building contents, evaluating, discussing etc. via the internet and social network. [6], [10], [2], [12]

**2.3. Active and Collaborative Learning Model for the Introductory Physics**

**2.3.1. Preparation for students take part in ACLM**

In the welcome lecture of the introductory physics, we surveyed 1$^{st}$ year students (K63) of the Faculty of Technology Education with voting paper. The summary table with 165 student's vote showing here:

| Learning supported variations in school | Frequency uses | Efficiently self-evaluation |
|---|---|---|
| - Email: 25 (15.15%) <br> - Chat: 14 (8.48%) <br> - Forum: 32 (19.39%) | - Rarely: 16 (9.7%) <br> - Sometimes: 92 (55.76%) | - Useless: 0 (0%) <br> - Not efficient: 5 (3.03%) <br> - Neutral: 28 (16.97%) |
| - Blog: 36 (21.82%)<br>- Social network: 108 (65.45%)<br>- YouTube: 45 (27.27%)<br>- Learning management system: 48 (29.09%)<br>- Others: 22 (13.33%) | - Frequently: 35 (21.21%) | - Less efficient: 97 (58.79%)<br>- Very efficient: 13 (7.88%) |
| Learning activities in school | Frequency uses | Efficiently self-evaluation |
| - Discussion: 153 (92.73%)<br>- Project: 5 (3.03%)<br>- Playing role: 7 (4.24%)<br>- Others: 4 (2.42%) | - Rarely: 24 (14.55%)<br>- Sometimes: 94 (56.97%)<br>- Frequently: 31 (18.79%) | - Useless: 1 (0.61%)<br>- Not efficient: 5 (3.03%)<br>- Neutral: 23 (13.94%)<br>- Less efficient: 103 (62.42%)<br>- Very efficient: 13 (7.88%) |

In the survey, students can choose multiple choices with the question for learning supported variations and learning activities in school. In spite of robustness, the result reveals experiences and ideas of students, which help instructor to prepare for student taking part in ACLM. According to the survey, activities with ICT support and enhanced learning activate solutions got highly appreciated by students. However the frequency uses is low because it is likely caused by the limit of learning space, a time of learning activities in the classroom, students ICT skills and ICT condition.

**2.3.2. Deploying ACLM with ICT support**

A/ Management of learning activities: In [5], authors suggest 4 phases in ACLM process: preparation → learning activities → discussion → summary. The details of activity depend on the learning condition and learning content. We go further in the Introductory Physics as follow:

Phase 1: preparation. Due to the problematic requirement, student study individual, in pair or small group. Instructor annouces tasks, guides the way to study. If working in a pair or group, students negotiate each duty. The number of members in a learning group should be chosen as 3-5 due to a reasonable duty arrangement, in our point of view. Working process must be presented by the instructor.

Phase 2: learning activities. Student actively self-studies given problem, searches documents, etc. He/she works individually to solve own problem.

Phase 3: discussion. Students discuss in pair or own group to make an addition, correct the project outcome. Writing down questions about own concerned subject need to be answered by the instructor.

Phase 4: summary. Students present and defend the result chosen by one of the mode:

+ A group-pair mode: two groups working in pair. Each must give criticizing and appreciating of partners result. They also ask and answer about the result of own and partners.

+ A class-presenting mode: individual or group presenting and defending.

Finally, instructor summarizes, gives added information, answers question, corrects the results of individual, duet or groups.

B/ Learning content design: Method of pattern language is taken to design learning content. [8] Classification and purpose uses of the pattern language method were presented in [7]. In the introductory physics course, temporary sample is applied due to the consistency with constructivism. The student is required to rebuild content according to the structure as follows:

1/ Group Name
2/ Member Name
3/ Name of Subject
4/ Comic: Graph related to the topic
5/ Context: A sample related to the field
6/ Target and Purpose: Target and purpose using a subject
7/ Features and applying condition: features and applying a condition of the subject
8/ Applying: activities, samples of subject application in life, technology, etc.
9/ Extra information: Addition about subject or suggestion finding more information with related subjects

With the time limitation of activities in class, the student is required to write down own duty result on A4 paper consistent with the above structure. The groups result is combined of members' achievements.

C/ Learning evaluation: formative and summative evaluation both used. At the conclusion of the course, students must participate in the final exam according to the training required to obtain a final score. A portion of an exam of theory question follows the above structure.

A middle mark of the student got by weighted average mark of individual and group activities. Activities include solving problems, making project, doing exercise, doing a middle exam, etc. A reasonable weighted mark was decided by the instructor and announced in class.

Evaluation points of individual and group are published. Members of a group negotiate evaluating weight point due to each contribution. A student's mark for a group solved problem is equal to the weight point multiplied with group's mark.

Group's mark is evaluated according to the result as follows:

| Idea (10p) | Oral presentation (10p) |
|---|---|
| Exact (10p) | Team support (10p) |
| Application (10p) | Answer (10p) |
| Presentation on paper… (10p) | Question (10p) |

Group's result is evaluated by the partner group in the group - pair mode or whole class in the class-presenting mode.

D/ ICT support: The efficiently ICT support is clearly shown in building rich multimedia content, searching, collaborated process, presenting etc. We have deployed the Learning Management System (LMS) of HNUE to make the online introductory physics course to support a student at the address: http://lms.hnue.edu.vn) with the mode of "Blended Learning". [3]

The course contains required subjects for individual or groups as task or assessment. The student can add submission via LMS. Presentation as a hard copy was shown as photo. All of requirements, evaluation guide, group working guide, etc. were published. A forum for the whole course and each subject was created and ready for discussing topics related and concerned. Extra information also was uploaded as attached documents or from other resources as videos for physics experiments, technology and website links related etc. The LMS system supports conference room for online discussing between students and instructors at a timely decision.

For efficient learning, instructor guides, student exploiting other ICT services as sharing documents, building contents etc via internet with the help of Google Drive, Drop box… and social networks as Facebook to get information and support quickly and in time because as the survey result shown students pay a great attention to communicate with social networks.

**2.3.3. Several problems in deploying ALCM and suggestion**

A/ Deploying result: we have deployed ALCM for part 1 of the introductory physics and surveyed students at the end of the course. With 176 responded votes, the result as follows:

Students' attitude to ALCM:

| Surveyed Item | Dislike/ Disagree | Neutral | Like/ Agree |
|---|---|---|---|
| Like methodology | 29 | 71 | 76 |
| Active learning improved | 27 | 53 | 96 |
| Self-studying improved | 22 | 41 | 111 |
| Efficiency | 61 | 68 | 67 |
| Group-partner mode suitable | 20 | 48 | 108 |
| Class-presenting mode suitable | 19 | 33 | 118 |
| Social network improved learning efficiency | 26 | 38 | 110 |
| LMS improved learning efficiency | 26 | 66 | 79 |
| Art (poetry, music…) made learning interest | 10 | 19 | 144 |

The survey result shows ACLM preferred by most of students who agree with learning methodology. However the efficiency of LMS voted at low rate, whereas a neutral vote at high was caused by the limitation of LMS-using skills and chit-chat liked psychology via social network of students. Art performance of students (playing poetry, music…) in the rest time or at the welcome session to reduce stress and enhance learning interest is one of the solutions mentioned in our another study. [9] This was supported by almost students.

We also asked students about the grounds of their spirit and the result showing:

| Reason in favour: | | Reason in unfavour: | |
|---|---|---|---|
| Own preparation allowance | 75 | Discuss disliked | 7 |
| Learning content related to life | 62 | Uncomfortable in presentation | 14 |
| Discussion with instructor allowance | 87 | Overload learning content | 16 |
| Doing many exercises | 10 | Lack of learning information | 58 |
| Knowledge extension | 57 | Lack of information searching skill | 43 |
| Working in pairs or group | 110 | Learning content unrelated to life | 14 |
| Presentation allowance | 90 | Boring methodology | 19 |
| Defense allowance | 86 | Group mode unsuitable | 13 |
| Question allowance | 91 | | |

Thus the reason for students' interest in ACLM was confirmed by the active and collaborative learning and developmental skills with activities which is essential for the future job. Their difficulties in learning caused by lack of skills in working and searching with documents and learning materials. Some students have not adapted with new learning methodology and relied on others. Some of them felt ashamed to make a presentation. The disadvantage of psychology should overcome with practice.

B/ Suggestion: ACLM requires the timely support of instructor to students. The student should be trained to have skills of self-solving problem and collaboration in learning. To meet such requirements, the facilities of ICT play important roles.

The LMS of HNUE is based on a well known open-source named Moodle, which supplies free services for courseware. It is necessary to research and invest to improve friendly facilities of LMS as registration, using resources for both student and instructor. If an application form connected with social network studied, the information of courseware would be transferred to the student and instructor timely which enhance the efficiency of learning.

## 3. Conclusion

Applying ACLM in the introductory physics course shows the advantage in competency-based education. Using ACLM in specific subjects training makes efficient in learning because students get benefit from active and collaborative working in which many soft skills for future job development. Deploying ICT in combination of social network and LMS with blended-learning mode will increase the efficiency of ACLM so it need to be studied and supported.

Acknowledgment: Author thanks to the support of the project SPHN-13-314.

# XÂY DỰNG MÔI TRƯỜNG HỌC TẬP CHỦ ĐỘNG VÀ HỢP TÁC TRONG KHÓA HỌC VẬT LÝ ĐẠI CƯƠNG CỦA KHOA SƯ PHẠM KỸ THUẬT

*Nguyễn Hoài Nam, Tiến sĩ, namnh@hnue.edu.vn,– khoa Sư phạm Kỹ thuật, Trường Đại học Sư phạm Hà Nội, Hà Nội, Việt Nam*

**Tóm tắt**. Mô hình học tập chủ động hợp tác được áp dụng trong đào tạo chuyên ngành cho thấy những ưu việt do vừa trang bị được kiến thức, tăng tính chủ động và rèn luyện nhiều kỹ năng của người học, giúp cho việc phát triển nghề nghiệp tương lai. Bài báo nghiên cứu và áp dụng mô hình học tập chủ động hợp tác để xây dựng môi trường học tập môn Vật lý đại cương với việc khai thác sự hỗ trợ của CNTT, đặc biệt là hệ thống mạng xã hội và eLearning của trường ĐHSP Hà Nội.

***Từ khóa:*** Sư phạm, Học tập chủ động và hợp tác, Vật lý đại cương, Học tập kết hợp, Môi trường học tập